\documentclass{article}
\usepackage{arxiv}
\usepackage[utf8]{inputenc} % allow utf-8 input
\usepackage[T1]{fontenc}    % use 8-bit T1 fonts
\usepackage{hyperref}       % hyperlinks
\usepackage{url}            % simple URL typesetting
\usepackage{booktabs}       % professional-quality tables
\usepackage{amsfonts}       % blackboard math symbols
\usepackage{nicefrac}       % compact symbols for 1/2, etc.
\usepackage{microtype}      % microtypography
\usepackage{lipsum}
\usepackage{graphicx}
\usepackage{amsmath}
\graphicspath{ {./} }
\title{iTARGET: Interpretable Tailored Age Regression for Grouped Epigenetic Traits}

\author{
  Zipeng Wu \\
  School of Mathematics\\
  University of Birmingham\\
  Birmingham, B15 2TT, UK \\
  \texttt{zxw365@bham.ac.uk} \\
  \And
  Daniel Herring \\
  School of Mathematics\\
  University of Birmingham\\
  Birmingham, B15 2TT, UK \\
  \texttt{d.g.herring@bham.ac.uk} \\
  \And
  Fabian Spill \\
  School of Mathematics\\
  University of Birmingham\\
  Birmingham, B15 2TT, UK \\
  \texttt{f.spill@bham.ac.uk} \\
  \And
  James Andrews\thanks{*Corresponding author} \\
  School of Mathematics\\
  University of Birmingham\\
  Birmingham, B15 2TT, UK \\
  \texttt{j.w.andrews@bham.ac.uk} \\
}

\begin{document}
\maketitle
\begin{abstract}
Accurately predicting chronological age from DNA methylation patterns is crucial for advancing biological age estimation. However, this task is made challenging by Epigenetic Correlation Drift (ECD) and Heterogeneity Among CpGs (HAC), which reflect the dynamic relationship between methylation and age across different life stages. To address these issues, we propose a novel two-phase algorithm. The first phase employs similarity searching to cluster methylation profiles by age group, while the second phase uses Explainable Boosting Machines (EBM) for precise, group-specific prediction. Our method not only improves prediction accuracy but also reveals key age-related CpG sites, detects age-specific changes in aging rates, and identifies pairwise interactions between CpG sites. Experimental results show that our approach outperforms traditional epigenetic clocks and machine learning models, offering a more accurate and interpretable solution for biological age estimation with significant implications for aging research.
\end{abstract}

% keywords can be removed
%\keywords{First keyword \and Second keyword \and More}

\section{Introduction}

DNA methylation is the process of chemically modifying DNA by adding methyl groups to CpG sites, which provides cells with an epigenetic code that dictates their function. The state of the methylation sites alter the expression of genes. Strikingly, the methylation pattern changes in aging. This observation led to the development of aging clocks - algorithms to estimate an individual's chronological \cite{AgingClocksChallenges} - from methylation patterns. Analyzing these methylation changes through advanced statistical or machine learning models makes achieving highly accurate age estimations possible. This technique is paramount in aging research\cite{healthyaging}.

%However, modeling DNA methylation age prediction effectively is a complex task due to two intrinsic properties of CpG methylation. Firstly, the correlation between CpG site methylation levels and age is not static but changes over time, which called epigenetic correlation drift in the paper. For instance, a CpG site might exhibit a high correlation with age in the 0-20 age range but a low correlation in the 20-30 age range. Consequently, relying on a static overall linear correlation value or weight can overlook CpG sites that are highly correlated with age in specific ranges, such as 40-50, but exhibit low correlation outside these ranges.

However, modeling DNA methylation age prediction effectively is complex due to two intrinsic properties of CpG methylation. Firstly, the correlation between CpG site methylation levels and age is not static but changes with age, which we call Epigenetic Correlation Drift (ECD). Specifically, DNA methylation levels change rapidly during early development and adolescence, stabilize during adulthood, and may alter again in older ages \cite{okada2023NonLinear, carlsen2023nonlinear2}. This nonlinearity\cite{NonlinearAging} poses significant challenges for traditional age prediction models, which often assume a uniform rate of change across the lifespan. Consequently, relying on a static overall linear correlation value can overlook CpG sites that are highly correlated with age in specific ranges, such as 40-50 years, but exhibit low correlation outside these ranges. Different CpG sites exhibit distinct methylation change patterns, called Heterogeneity Among CpGs (HAC). For example, a CpG site might have a high correlation with age during one age stage but a low correlation in the next, while other CpG sites maintain a high correlation in the subsequent stage. 

Moreover, modern technologies enable one to measure the methylation state of 450,000 or more CpG sites. Hence, selecting important CpG sites for modelling is one of the crucial steps.  
All related work typically employs linear regression along with Pearson and Spearman coefficients to select CpG sites\cite{belsky2022DunedinPACE,hannum2013genome,horvath2epigenetic2018,HRSInCHPhenoAge,varshavsky2023accurate,lin2016dna,PhenoAge,YingCausAgeDamAgeAdaptAge}, operating under the assumption of a linear relationship or monotonicity. However, this approach faces significant drawbacks due to a need for more consideration of the ECD. The correlation between CpG site methylation levels and age is not static but varies over time. This variability means that modelling based on this kind of CpG site selection may fail to accurately capture the age-related methylation patterns across all age ranges. For instance, a CpG site may exhibit a high correlation with age in the 0-20 age range and a low correlation in the 20-30 age range. Consequently, using a single linear correlation value or weight for the entire age range may overlook CpG sites highly correlated with age only within specific intervals, such as 40-50 years, but show low correlation outside these intervals.  

While machine learning and deep learning techniques have shown significant success in related works on biological age prediction\cite{varshavsky2023accurate,DL2022pan,galkin2021deepLmage,deeplNN2021age,levy2020methylnetdeepl,vidaki2017dnaDeepl}, the need for model interpretability has become increasingly important in this domain. The ability to explain model predictions is crucial for understanding the underlying aging mechanisms, especially in clinical settings where interpretability can guide informed decision-making. \cite{qiu2023EAIIMLAgeClock} introduces the ENABL Age framework, which combines gradient-boosted trees (GBM) with Shapley Additive Explanations (SHAP), a type of interpretable machine learning technique. This approach not only achieves high predictive accuracy but also provides individualized explanations for each prediction, offering deeper insights into the contributing factors of biological age. Despite its promise, the application of interpretable machine learning methods in this area remains limited, with this study being one of the few that address the need for transparency in age prediction models.

To address these limitations, we firstly propose a novel two-phase algorithm for DNA methylation age prediction that explicitly accounts for the nonlinear nature of methylation changes across different age phases. Our approach first uses Facebook AI Similarity Search (FAISS) to cluster methylation profiles into age groups based on their similarity\cite{douze2024faiss}. Rapid clustering can be done without dimensionality reduction due to the ability of FAISS to efficiently handle ultra-high dimensional data. The second phase involves training a specific model for each age group. Secondly, we select different DNA CpG sites for different age groups for training the Explainable Boosting Machine (EBM)\cite{EBM2022interpretability,lou2013EBM,caruana2015EBM} model, which not only improves the precision
of age predictions but also enhances the interpretability
and scalability of the models. By focusing on specific age
groups, we reduce the variability within each group, leading
to more reliable and interpretable results.

%Our research aims to fill the gap left by traditional methods by providing a more nuanced and accurate prediction of biological age through the use of phase-specific models. The proposed two-phase algorithm not only improves the precision of age predictions but also enhances the interpretability and scalability of the models. By focusing on specific age groups, we reduce the variability within each group, leading to more reliable and interpretable results.
The main contributions of this study can be summarised as follows:
\begin{itemize}
    \item \textbf{Identification of Decade-Specific Age-Related Biomarkers:} Calculation the CpG sites with the strongest linear correlations within each decade-specific age window, allows more accurate identification of the key age-related biomarkers at different stages of aging, and addresses the variability and nonlinear relationships across the broader age span. 
    \item \textbf{Detection of Synergistic Interactions Between CpG Sites:} Explainable Boosting Machines (EBM) enable global and local insights into the influence of CpG sites on the overall model and individual predictions. Additionally, the detection and modelling of interactions between features, offers a valuable tool for detecting synergistic interactions between CpG sites.

    \item \textbf{Novel Two-Stage DNA Methylation Age Prediction:} Development of a two-stage method that first uses Pearson correlation for adaptive age group-specific CpG site selection, that addresses CpG heterogeneity, and then employs a FAISS-based age group identification followed by precise prediction allows for mitigation of Epigenetic Correlation Drift.
    
    \item Experimental results demonstrate that the proposed method outperforms traditional epigenetic clocks and other interpretable machine learning prediction models in terms of accuracy.  
\end{itemize}

\section{Methodology}
\begin{figure*}[!t] 
    \centering
    \includegraphics[width=\textwidth]{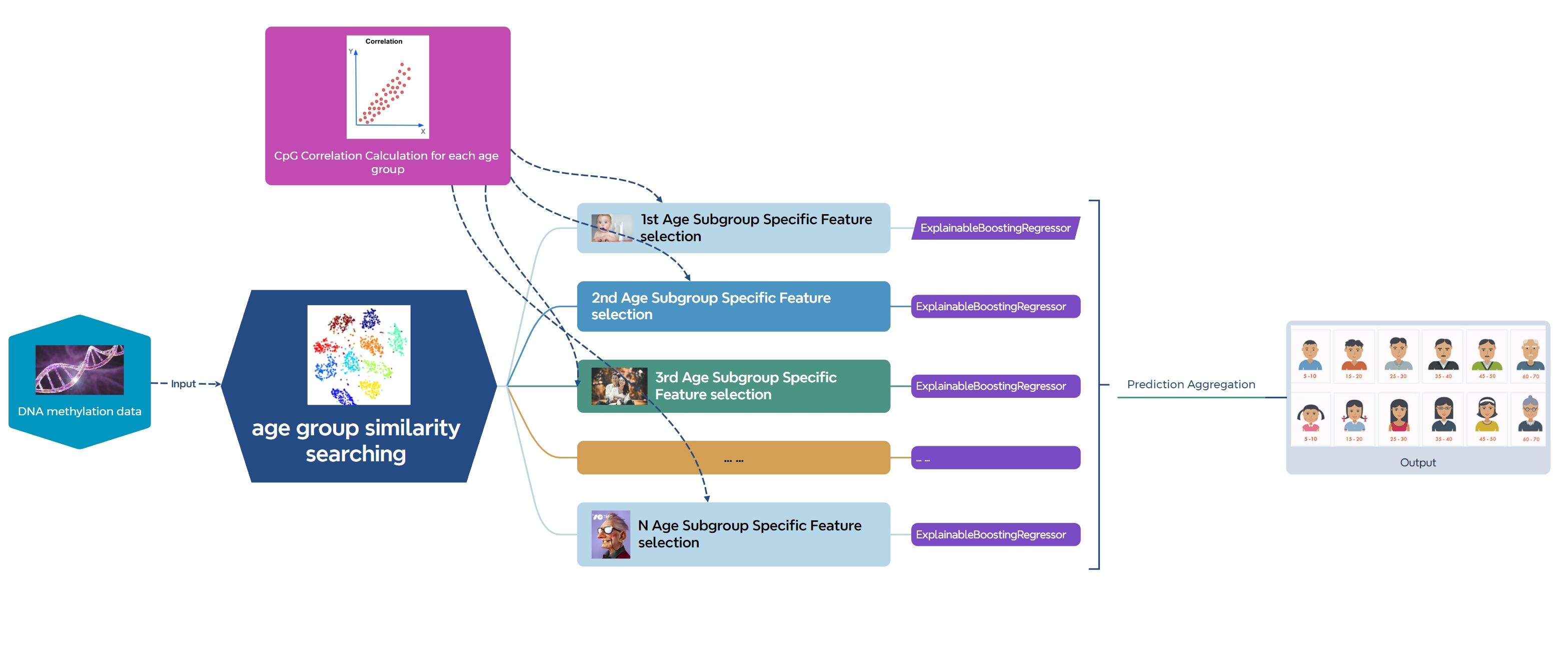}
    \caption{Overview of our proposed method.}
    \label{fig:our_methods}
\end{figure*}
Our proposed approach addresses two primary challenges in DNA methylation age prediction: Epigenetic Correlation Drift and CpG site heterogeneity. Epigenetic Correlation Drift refers to the temporal variability in the correlation between DNA methylation levels at specific CpG sites and chronological age, indicating that the relationship between methylation and age varies across different age ranges. Additionally, CpG site heterogeneity reflects the diverse patterns of methylation changes, necessitating a more refined feature selection process.

Let $X$ represent the DNA methylation data matrix, where $X$ is an $N \times M$ matrix, with $N$ representing the number of samples and $M$ the total number of CpG sites. The vector $y$, which corresponds to the chronological ages of the $N$ samples, is an $N \times 1$ vector.

We employ two age grouping strategies. The first divides the age range into decade-sized intervals: $[0-10),[10-20),[20-30), \ldots,[90-100),[100+ $. This approach is motivated by its interpretability, as decade intervals are commonly used and easily understood, making the results accessible to a broad audience. The second grouping is based on research by \cite{lehallier2019undulating}, which identified key inflection points in aging at approximately 34,60, and 78 years. This strategy divides the age range into four segments: [0-34), [34-60), [60-78), and 78+, aligning with significant biological and proteomic changes that correspond to shifts in aging patterns.

Within each age group, we calculate Pearson correlation coefficients for DNA methylation data to identify CpG sites most strongly correlated with age. This allows us to capture localized, linear relationships within each age range, potentially identifying age-specific biomarkers.

\subsection{Age Group-Specific CpG Correlation Calculation and Feature Selection}

We construct an age-specific matrix for each age group and compute Pearson correlation coefficients for each CpG site within the matrix. For each age group $g_i$ :
$$
X_{g_i}=\left[\begin{array}{cccc}
x_{11}^{\left(g_i\right)} & x_{12}^{\left(g_i\right)} & \cdots & x_{1 M}^{\left(g_i\right)} \\
x_{21}^{\left(g_i\right)} & x_{22}^{\left(g_i\right)} & \cdots & x_{2 M}^{\left(g_i\right)} \\
\vdots & \vdots & \ddots & \vdots \\
x_{N_{g_i} 1}^{\left(g_i\right)} & x_{N_{g_i} 2}^{\left(g_i\right)} & \cdots & x_{N_{g_i} M}^{\left(g_i\right)}
\end{array}\right]
$$
where $X_{g_i}$ is an $N_{g_i} \times M$ matrix, $N_{g_i}$ is the number of samples in age group $g_i$.

The Pearson correlation coefficient $\rho_j^{\left(g_i\right)}$ for CpG site $j$ in age group $g_i$ is given by:
$$
\rho_j^{\left(g_i\right)}=\frac{\sum_{k=1}^{N_{g_i}}\left(x_{k j}^{\left(g_i\right)}-\bar{x}_j^{\left(g_i\right)}\right)\left(y_k^{\left(g_i\right)}-\bar{y}^{\left(g_i\right)}\right)}{\sqrt{\sum_{k=1}^{N_{g_i}}\left(x_{k j}^{\left(g_i\right)}-\bar{x}_j^{\left(g_i\right)}\right)^2} \sqrt{\sum_{k=1}^{N_{g_i}}\left(y_k^{\left(g_i\right)}-\bar{y}^{\left(g_i\right)}\right)^2}}
$$
where: $x_{k j}^{\left(g_i\right)}$ is the methylation level at CpG site $j$ for sample $k$ in age group $g_i$; $y_k^{\left(g_i\right)}$ is the age of sample $k$ in age group $g_i$; $\bar{x}_j^{\left(g_i\right)}$ is the mean methylation level at CpG site $j$ in age group $g_i$ and $\bar{y}^{\left(g_i\right)}$ is the mean age in age group $g_i$.

By focusing on smaller age ranges, we can capture CpG sites exhibiting relative easy relationships with age, thus addressing age-related variability and nonlinear relationships across the broader age span. Based on these correlation coefficients, we select the top 30 CpG sites for each age group, resulting in a tailored feature set that is more relevant for age prediction within that group.
The top 30 CpG sites for each age group are selected based on the largest correlation values, 
$\text{max}(\left|\rho_j^{\left(g_i\right)}\right|)$ :
$$
X_{g_i}^{\text {selected }}=\left[\begin{array}{cccc}
x_{11}^{\left(g_i\right)} & x_{12}^{\left(g_i\right)} & \cdots & x_{1,30}^{\left(g_i\right)} \\
x_{21}^{\left(g_i\right)} & x_{22}^{\left(g_i\right)} & \cdots & x_{2,30}^{\left(g_i\right)} \\
\vdots & \vdots & \ddots & \vdots \\
x_{N_{g_i} 1}^{\left(g_i\right)} & x_{N_{g_i} 2}^{\left(g_i\right)} & \cdots & x_{N_{g_i}, 30}^{\left(g_i\right)}
\end{array}\right]
$$
where $X_{g_i}^{\text {selected }}$ is an $N_{g_i} \times 30$ matrix.

We chose to focus on only 30 CpG sites because this selection strikes an optimal balance between interpretability and predictive performance. This decision is motivated by the findings of \cite{varshavsky2023accurate}, whose experiments demonstrated that even with just 30 CpG sites, it is possible to achieve strong predictive accuracy.

\subsection{Age Group-Specific Model Training}

Using the top 30 features identified for each age group, we train specific models tailored to each age group. 
A regression model $f_{g_i}$ is trained for each age group $g_i$ using $X_{g_i}^{\text {selected }}$ and $y_{g_i}$ : $\hat{y}_{g_i}=f_{g_i}\left(X_{g_i}^{\text {selected }}\right)$.
This strategy reduces the complexity of the relationships within each group, making linear assumptions more valid and improving the overall accuracy of the predictions. By having tailored feature sets for each age group, our models can better capture the unique methylation patterns associated with different stages of aging.

\subsubsection{Explainable Boosting Machine}

Explainable Boosting Machine (EBM)\cite{lou2013EBM,caruana2015EBM,EBM2022interpretability} is a generalized additive model (GAM) that incorporates boosting to improve accuracy while maintaining interpretability. The EBM model can be expressed symbolically as follows:
$$
h(E[\hat{y}])=\beta_0+\sum_{j=1}^M f_j\left(x_j\right)
$$
where $h$ is the link function that adapts the model to different settings, such as regression or classification, $E[\hat{y}]$ is the expected value of the prediction, $\beta_0$ is the intercept and $f_j\left(x_j\right)$ represents the learned feature function for the $j$-th feature $x_j$.

To improve accuracy, EBM can automatically detect and include pairwise interaction terms between features:
$$
h(E[\hat{y}])=\beta_0+\sum_{j=1}^M f_j\left(x_j\right)+\sum_{r=1}^M \sum_{s=r+1}^M f_{r s}\left(x_r, x_s\right)
$$
where $f_{r s}\left(x_r, x_s\right)$ represents the interaction function between the $r$-th and $s$-th features. The pairwise interaction terms allow the model to capture interactions between features while retaining the overall interpretability of the model. 
For regression tasks, the link function $h$ is typically the identity function, so the EBM model becomes:
$$
\hat{y}=\beta_0+\sum_{j=1}^M f_j\left(x_j\right)+\sum_{r=1}^M \sum_{s=r+1}^M f_{r s}\left(x_r, x_s\right)
$$
where $\hat{y}$ is the predicted age based on the DNA methylation profile.

In the presence of both individual and interaction effects, the total contribution from a feature $x_j$ (including any interactions it participates in) is the sum of the individual contribution and the contributions from all interaction terms involving that feature:
Total Contribution of $x_j=f_j\left(x_j\right)+\sum_{s \neq j} f_{j s}\left(x_j, x_s\right)$
where: $f_j\left(x_j\right)$ is the direct contribution from feature $x_j$ and $f_{j s}\left(x_j, x_s\right)$ are the interaction terms involving feature $x_j$ and any other feature $x_s$.

\subsection{Similarity Searching Age Group identification}
In the initial stage of age prediction, we employ similarity searching techniques to accurately identify the relevant age group for each sample. We utilize FAISS (Facebook AI Similarity Search) \cite{douze2024faiss}, a powerful library optimized for searching similar vectors in high-dimensional spaces, making it particularly well-suited for our DNA methylation data. Given the vast dimensionality of the data (over 450,000 CpG sites in the Infinium Human Methylation 450K BeadChip and more than 930,000 CpG sites in the Infinium MethylationEPIC v2.0 array), FAISS efficiently handles this complexity, ensuring rapid and precise age group classification. The goal is to find the age group $g_i$ whose samples' methylation profiles are most similar to $x_{\text {new }}$.

The FAISS index $\mathcal{I}$ is initialized using the L2 (Euclidean) distance with dimensionality $d=M$. Training data $X_{\text {train }}$ is then added to this index, constructing $\mathcal{I}=$ FAISSIndex $\left(X_{\text {train }}\right)$. For each test sample in $X_{\text {test }}$, a similarity search is performed to find the $k$ nearest neighbors, yielding the distance matrix $\mathcal{D}$ and the indices $\mathcal{I}_k=\operatorname{FAISS}\left(X_{\text {test }}, \mathcal{I}, k\right)$. The predicted age group $\hat{y}_{\text {test }, i}$ for the $i$-th test sample is determined by the age group of the nearest neighbor.
The classification performance in this step has an impact on the performance of the final prediction, and we set up comparison experiments.

\subsection{Age Group Specific model Accurate Age Prediction}
Once the relevant age group is identified for a sample, the age-group-specific model makes the final age prediction, ensuring accurate and precise age estimation. The model uses the identified age group $g_i^*$ and the corresponding regression function $f_{g_i^*}$ to make prediction:
$$
\hat{y}_{\text {test }, i}=f_{g_i^*}\left(x_{\text {test }, i}^{\text {selected }}\right)
$$

Our methodology focuses on localized relationships and reduces complexity to enhance the predictive accuracy of DNA methylation age prediction models. The tailored feature sets for each age group and the two-stage prediction process address the inherent challenges posed by Epigenetic Correlation Drift and CpG site heterogeneity, making our approach robust and reliable for age estimation across different age ranges.

\section{Comparison of Prediction Accuracy of Epigenetic Clocks}

To verify the performance of our proposed method, we design three groups of experiments. The primary aim is to compare the proposed approach against traditional established epigenetic clocks and classic linear regression models and to assess the impact of our novel approach to data segemetation by interpretable age grouping.

The first group of experiments involved comparing existing pre-trained epigenetic clocks available in the Biolearn Python library, providing a comprehensive clock model suite. The baseline models include \emph{Horvathv1} \cite{horvath2013dna}, \emph{Hannum} \cite{hannum2013genome}, \emph{Lin} \cite{lin2016dna}, \emph{PhenoAge} \cite{PhenoAge}, \emph{YingCausAge} \cite{YingCausAgeDamAgeAdaptAge}, and \emph{Horvathv2} \cite{horvath2epigenetic2018}. These models predict chronological age or aging rate based on DNA methylation data from various tissues.

The second group of experiments compares the proposed approach with classic linear regression models trained on the same dataset. We select ElasticNet and Lasso Regression as our baseline models due to their established efficacy in this field. Inspired by Varshavsky et al. \cite{varshavsky2023accurate}, who used a Gaussian process regression, we incorporate a model predicting age using a compact set of 30 CpG sites identified through correlation analysis and clustering. This model demonstrates superior accuracy, with a median prediction error of 2.1 years on held-out blood samples. We also select the top 30 CpG sites with the highest Pearson correlation with age across the entire dataset for our experiments, and use Gaussian process regression as our baseline model.

The third set of experiments compares two age grouping strategies for DNA methylation age prediction. The first strategy uses decade-sized intervals (e.g., [0-10), [10-20), ..., [90-100)) for ease of interpretability. The second strategy, informed by \cite{lehallier2019undulating}, divides ages into segments at key inflection points: [0-34), [34-60), [60-78), and 78+, aligning with significant biological shifts observed in plasma proteome profiles.

All experiments were conducted using Python 3.9, with FAISS-CPU 1.8.0 for similarity search, Scikit-learn 1.5.0 for machine learning algorithms and biolearn 0.4.3 for traditional pretrained epigenetic clocks, and custom scripts for preprocessing and feature selection. All code is available on Github\footnote{https://github.com/WuzipengYL/iTARGET-Interpretable-Tailored-Age-Regression-for-Grouped-Epigenetic-Traits.}. The computations were performed on Intel I7 14700k processors and 32 GB RAM. To ensure the robustness of the results, we performed five cross-validations on the entire dataset.

All the experimental conditions as well as the code are at available on Github\footnote{https://github.com/WuzipengYL/iTARGET-Interpretable-Tailored-Age-Regression-for-Grouped-Epigenetic-Traits.}, which contains the top30 CpG sites with the maximum correlation for each age group, as well as a detailed explanation of the model for each age group.

\subsection{Dataset and Preprocessing}

For our experiments, we utilize a publicly available dataset of DNA methylation profiles, comprising 11,910 blood-derived methylomes from donors aged 0 to 103 years, as detailed in the study by Varshavsky et al. \cite{varshavsky2023accurate}. This dataset\footnote{ https://www.ncbi.nlm.nih.gov/geo/query/acc.cgi?acc=GSE207605.}, assembled from 19 genome-wide methylation array studies, includes data from the Illumina Infinium HumanMethylation450 BeadChip and MethylationEPIC BeadChip platforms.

The choice of this dataset is driven by its comprehensive coverage and representativeness. Such a large and diverse dataset ensures robustness and generalizability of our model across different age groups and technical platforms. The dataset is divided into training and testing sets, with 80\% (9,523 samples) used for training and 20\% (2,387 samples) reserved for testing. To ensure that there are samples in each age group, specifically, our train and test sets are divided by randomly selecting inside different age groups and aggregating them into the final test and training sets.

Preprocessing of the methylation data followed standard protocols, including quality control, normalization, and imputation of missing values. Specifically, potential age-related CpG sites were filtered down to 2,374 columns based on the criteria established by Varshavsky et al. \cite{varshavsky2023accurate}, ensuring the focus on relevant features. Any remaining missing values were imputed using the mean value method, ensuring consistency and completeness in the dataset.

\subsection{Baseline Models}
To ensure a comprehensive evaluation of our proposed method, we compare it against several established and widely recognized baseline models in the field of DNA methylation age prediction:

\begin{itemize}
    \item \textbf{Horvathv1 (2013)}: This multi-tissue age predictor uses 353 CpG sites and is widely regarded as a benchmark in the field of epigenetic aging clocks \cite{horvath2013dna}.
    \item \textbf{Hannum (2013)}: Focused on blood tissue, this model uses 71 CpG sites and was one of the first to demonstrate the feasibility of predicting age from DNA methylation data \cite{hannum2013genome}.
    \item \textbf{Lin (2016)}: This model refines predictions using 99 age-associated CpG sites and has shown a high correlation between predicted and chronological age in blood samples \cite{lin2016dna}.
    \item \textbf{PhenoAge (2018)}: Developed to predict phenotypic age, this model is based on 513 CpGs and is particularly noted for its ability to predict mortality and morbidity risk \cite{PhenoAge}.
    \item \textbf{Horvathv2 (2018)}: An updated version of the original Horvath clock, it integrates data from both skin and blood tissues \cite{horvath2epigenetic2018}.
    \item \textbf{ElasticNet}: ElasticNet is a Classic linear regression model.  It was selected to test the classic linear regression model trained by the same datasets and Pearson coefficient based feature selection \cite{ElasticNet}.
    \item \textbf{Gaussian Process Regression(2023)}: This model can capture the complex non-linear relationships within the data, which has been recently applied to age prediction tasks as Gp\_age\cite{varshavsky2023accurate}.
    \item \textbf{iTARGET-(34-60-78)}: This approach segments ages into biologically significant intervals: [0-34), [34-60), [60-78), and 78+.
    \item \textbf{iTARGET-ideal}: 
    Assuming perfect age group classification in Stage 1, each specimen is accurately assigned to its corresponding age-specific model for DNAm age calculation. This approach is well-suited for aging-related studies but is less applicable to forensic identification without a prior estimate of the age range.
\end{itemize}

\subsection{Results and Discussion}

%The table compares our proposed method's Mean Absolute Error (MAE) and Root Mean Squared Error (RMSE) against various established baseline models, including EBM, Linear Lasso, ElasticNet, and several traditional epigenetic clocks.

\begin{table}[!t] 
\centering
\caption{Comparison of MAE and RMSE for Different Methods}
\label{tab:1}
\begin{tabular}{cccc}
\hline 
Methods & Year & MAE & RMSE \\
\hline 
iTARGET-Decade & - & $\boldsymbol{3.7752}$ & 5.8164 \\
iTARGET-(34-60-78) & - & 3.9179 & 6.2646 \\
Linear Lasso & - & 4.4514 & $\boldsymbol{5.7082}$ \\
Gaussian Process Regression & 2023 & 4.8943 & 8.0822 \\
Linear ElasticNet & - & 5.9864 & 7.6171 \\
Hannum & 2013 & 6.6072 & 7.9169 \\
Horvathv2 & 2018 & 6.9330 & 8.3756 \\
Lin & 2016 & 9.0995 & 11.7827 \\
HRSInCHPhenoAge & 2022 & 9.7581 & 11.6568 \\
Horvath & 2013 & 13.2526 & 15.3902 \\
PhenoAge & 2018 & 15.8259 & 18.9178 \\
\hline
iTARGET-ideal  & - & 1.5744 & 2.5867 \\
\hline
\end{tabular}
\end{table}

Overall, the results of our experiments, summarized in Table \ref{tab:1}, demonstrate that our iTARGET-decade achieved the lowest Mean Absolute Error (MAE) among all evaluated models, with an MAE of 3.7752 in the decade-based grouping strategy. While Linear Lasso achieved a slightly lower RMSE of 5.7082 compared to iTARGET-decade's RMSE of 5.8164, our method still outperforms in terms of MAE, which is a crucial metric for accuracy in age prediction. This superior performance underscores the robustness of our age grouping strategy and targeted feature selection process. Importantly, our approach is entirely interpretable, selecting only the top 30 CpG sites for each age group, which provides clear insights into the biological relevance of each CpG site, making the model both accurate and easy to understand.

The "iTARGET-ideal " scenario represents an optimal condition where the correct age group classification is perfectly achieved. In this ideal scenario, iTARGET-ideal demonstrates exceptionally low error rates (MAE = 1.5744, RMSE = 2.5867), showcasing the potential for highly precise DNA methylation age prediction. This result underscores the strength of our approach when applied under ideal conditions, where each sample is accurately assigned to its corresponding age-specific model. This ideal scenario highlights that when age groups are correctly identified, iTARGET's focused selection of 30 CpG sites per group can lead to remarkably accurate predictions. Although achieving perfect classification in real-world applications may be challenging, this result serves as a benchmark, illustrating the theoretical upper limit of iTARGET's performance. It also suggests that further refinement in the classification step could bring practical applications closer to this ideal performance. In practical applications, particularly in contexts where the chronological age is known or can be estimated, iTARGET-ideal can be directly applied, leveraging the identified age group to enhance prediction accuracy. 

\subsubsection{Comparison with Pretrained Traditional Epigenetic Clocks}

Our proposed method demonstrates superior performance in DNA methylation age prediction compared to traditional pretrained epigenetic clocks, such as \emph{Horvathv1}, \emph{Hannum}, \emph{Lin}, \emph{PhenoAge}, and \emph{Horvathv2}. iTARGET-decade's MAE was 3.7752, and the RMSE was 5.8164, significantly lower than all the baseline models.

For instance, the \emph{Horvathv1} clock, a widely recognized benchmark, reported an MAE of 13.2526 and an RMSE of 15.3902. These values are higher than those of iTARGET-decade by 9.4774 in MAE and 9.5738 in RMSE, underscoring the effectiveness of our approach. The \emph{Hannum} model, another well-regarded predictor, recorded an MAE of 6.6072 and an RMSE of 7.9169, exceeding iTARGET-decade's errors by 2.832 in MAE and 2.1005 in RMSE. One reason for the poorer performance of these clocks is that they were trained on older datasets with varying sample sizes and data quality. While their pre-trained models, available in the Biolearn library, offer computational convenience, their generalizability may be limited. Additionally, some CpG sites used in these earlier models may have been removed or updated in newer platforms, such as the transition from EPIC v1.0 to the Infinium Human Methylation 450K BeadChip, further contributing to the differences in accuracy.
\subsubsection{Comparison with Machine Learning Models and Impact of Grouping Strategy}

iTARGET-decade also outperformed classic machine learning models such as Linear Lasso and ElasticNet. While Linear Lasso achieved an RMSE of 5.7082, slightly better than iTARGET-decade's RMSE, its higher MAE of 4.4514 indicates less precision in individual predictions. Gaussian Process Regression (GP), a method known for capturing nonlinear relationships, reported an MAE of 4.8943 and an RMSE of 8.0822. Despite its sophisticated approach, GP did not perform as well as Proposed approaches. This is likely because our age grouping strategy effectively simplifies the complexity of the relationships within each age segment, making the modeling process more straightforward and accurate.

The decade-based age grouping achieved a slightly lower MAE of 3.7752 compared to the 34-60-78 biologically significant inflection points grouping, which had an MAE of 3.9179, indicating more precise predictions across uniform ten-year intervals. This approach likely benefits from capturing localized, linear relationships in methylation patterns. Moreover, our proposed approaches identify the top 30 most important CpG sites within each age group, a novel contribution that enhances the understanding of key biomarkers across different life stages and offers valuable insights into the variations in aging rates throughout the lifespan. 
%This focused approach ensures that the model remains interpretable and manageable, offering clear explanations for the predictions it generates, which is often a limitation in more complex models that use a larger set of features.

\section{Identification of Decade-Specific Age-Related Biomarkers}

Understanding how DNA methylation patterns change across different stages of life is crucial for accurately predicting biological age and uncovering the underlying mechanisms of aging. Aging is not a uniform process; it varies across different periods of life, with some stages experiencing accelerated or decelerated aging. This variability can be captured by examining the contribution and interaction of CpG sites within specific age segments.

\begin{figure}[!t] 
    \centering
    \includegraphics[width=0.65\textwidth]{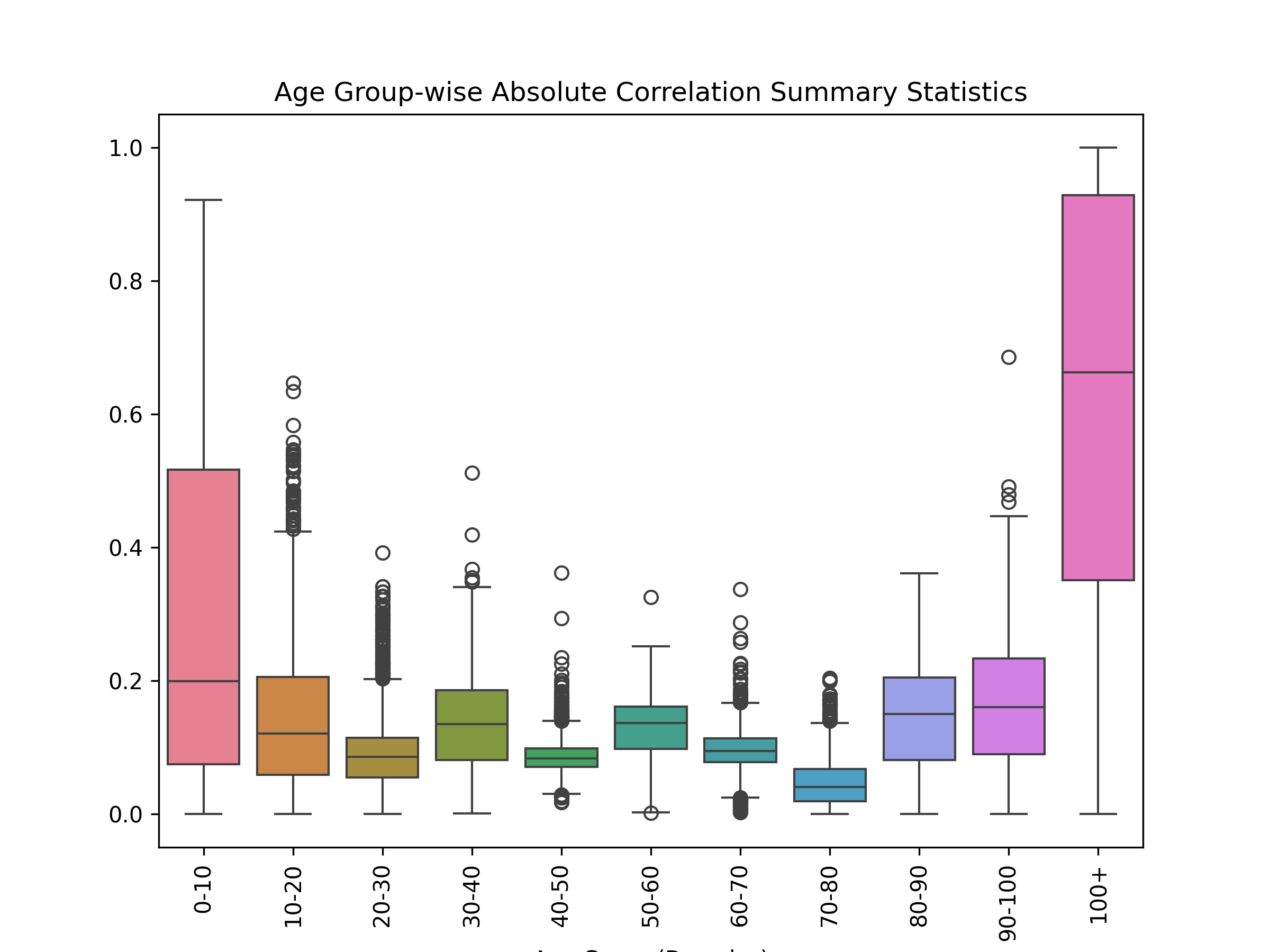}
    \caption{Box plot of absolute Pearson correlation coefficients of 2,374 age-related CpG sites across different decade windows.}
    \label{fig:correlation_summary}
\end{figure}

Figure \ref{fig:correlation_summary} presents a box plot of the absolute Pearson correlation coefficients between 2,374 age-related CpG sites and the corresponding age windows. The graph illustrates the distribution of overall methylation rates within each age group in relation to their correlation coefficients, revealing distinct differences between the age windows. Notably, age groups with correlation coefficients closer to zero tend to exhibit slower rates of aging.

In the 0-10 age range, the correlation is remarkably high, indicating a rapid rate of aging or growth during this period. A similarly fast rate of change is observed in the 10-20 age range. However, the aging rate appears to slow down in the 20-30 age range, and further deceleration is observed in the 40-50 and 70-80 age ranges. Interestingly, the 100+ age group shows a very high correlation, likely due to the smaller sample size and increased variability at the end of life. This high correlation suggests a faster rate of aging in this group, although this conclusion warrants validation with a larger sample size. These findings underscore the variability in aging rates across different life stages, with certain age windows exhibiting rapid changes in methylation patterns, while others reflect a slower pace of aging.

\begin{figure}[!t] 
    \centering
    \includegraphics[width=0.40\textwidth]{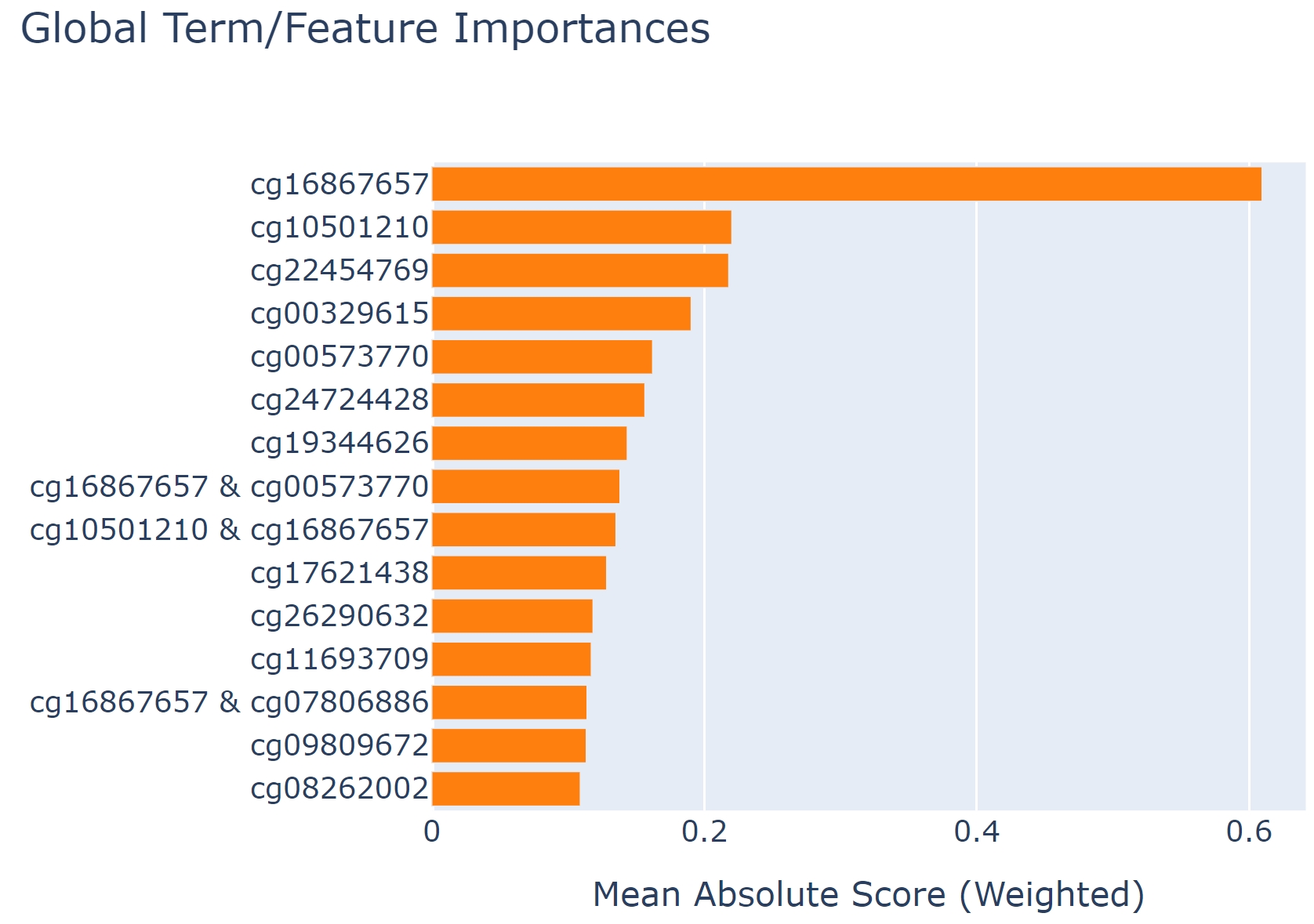}
    \includegraphics[width=0.40\textwidth]{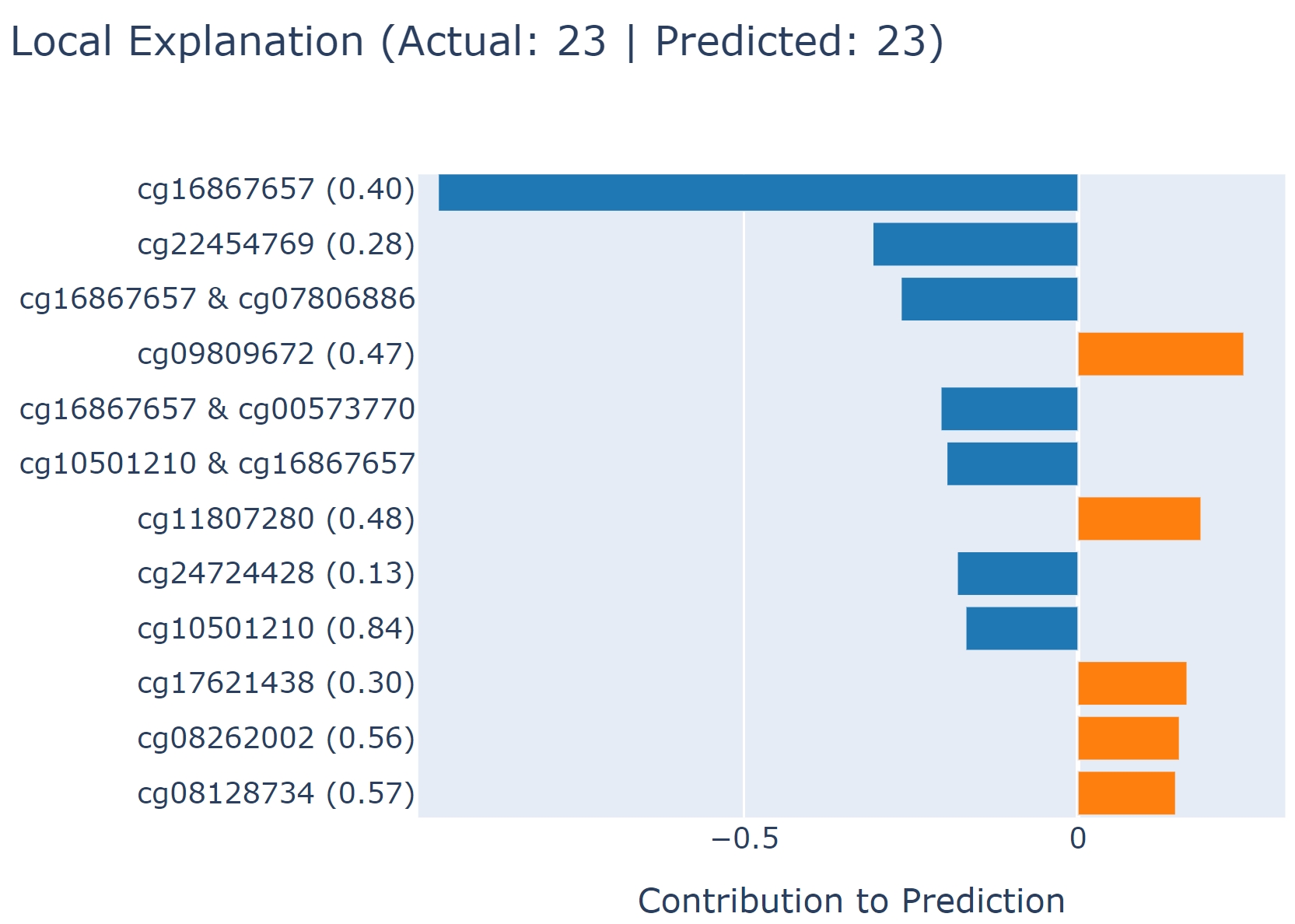}
    \caption{Global and local feature importance (contribution) for age group 20.0-30.0. The global feature importance highlights the CpG sites that have the most significant overall contribution to the age prediction model, while the local feature importance shows how these CpG sites contribute to individual predictions within this age group.}
    \label{fig:age_20_30_features}
\end{figure} 

\begin{figure}[!t] 
    \centering
    \includegraphics[width=0.45\textwidth]{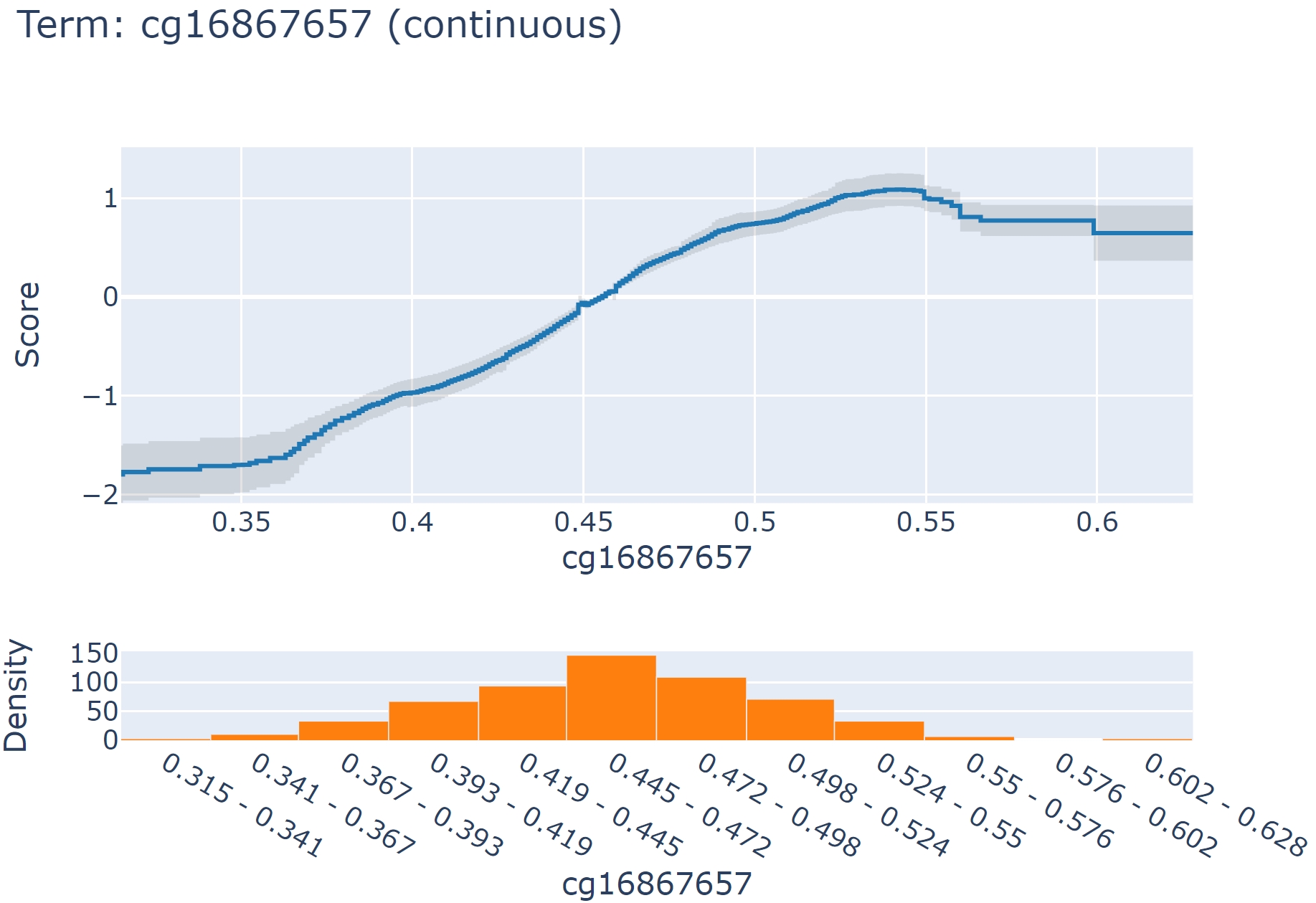}
    \includegraphics[width=0.45\textwidth]{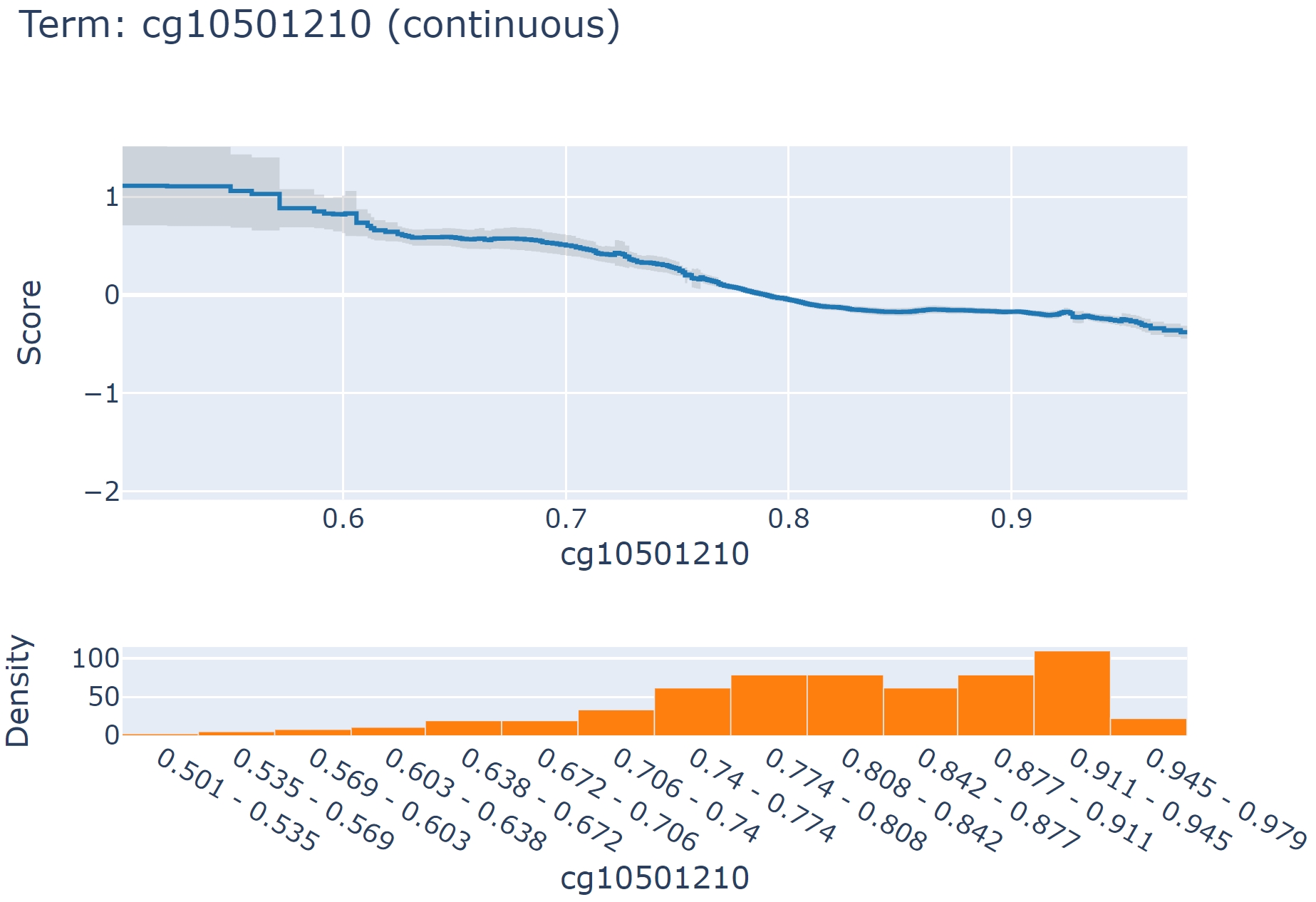}
    \caption{Top two CpG sites for age group 20.0-30.0.The top two CpG sites with the strongest influence on age prediction were identified which include graphs of the contribution of the two CpG sites to the prediction as a function of its value, and graphs of the distribution of the values of the two CpG sites.}
    \label{fig:age_20_29_cpg_top1_2}
\end{figure}

\begin{figure}[!t] 
    \centering
    \includegraphics[width=0.4\textwidth]{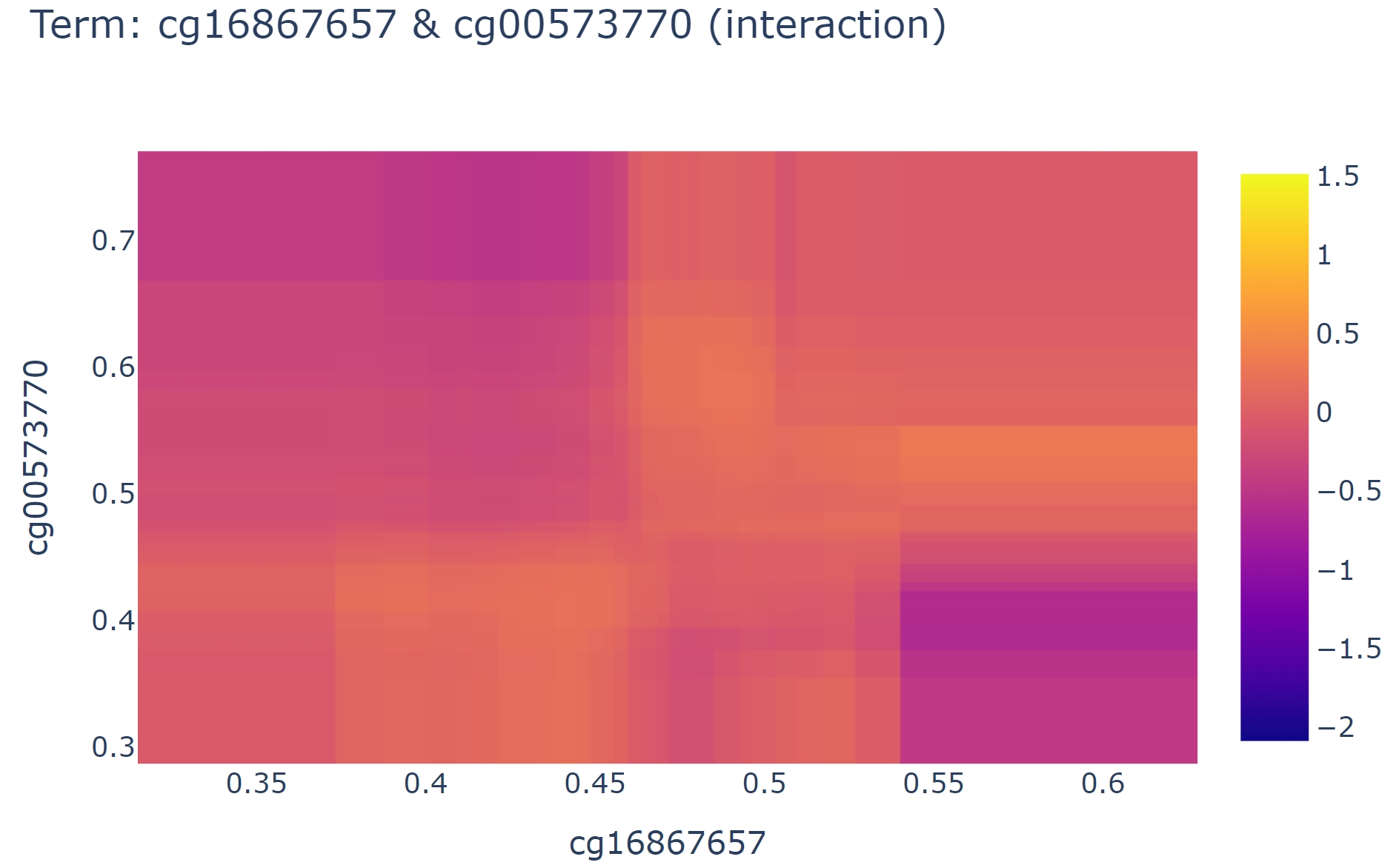}
    \includegraphics[width=0.4\textwidth]{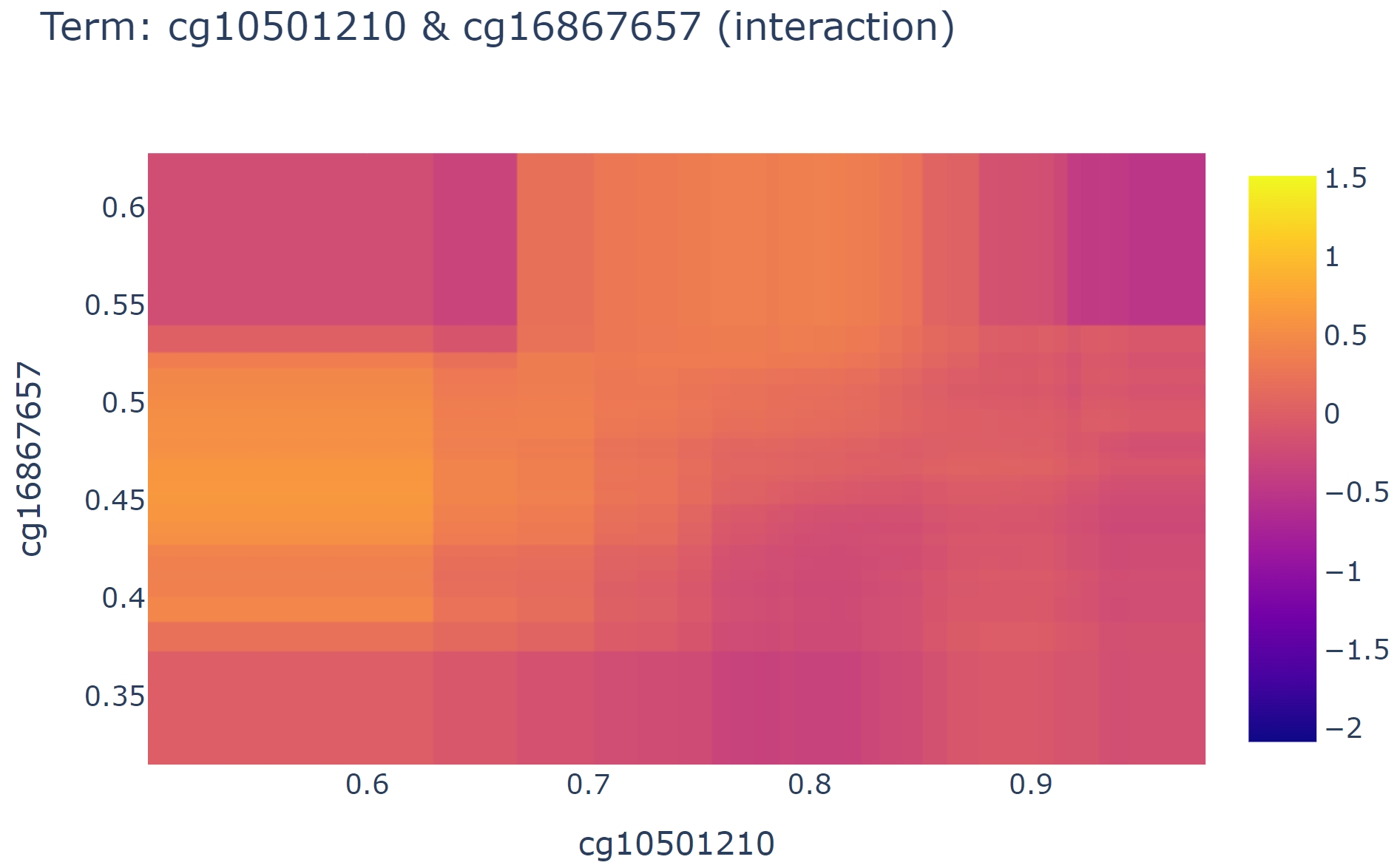}
    \caption{Heatmaps illustrating the interaction effects between CpG sites for the age group 20.0-30.0. The contribution of interactions between the two CpG sites to the predictive model are displayed, with colors closer to yellow indicating a stronger positive contribution and colors closer to purple indicating a stronger negative contribution. This visualization highlights the nuanced effects of CpG site interactions on age prediction within this specific age group.}
    \label{fig:age_20_29_cpg_interaction}
\end{figure}
For brevity, we chose the model for the 20-30-year-olds as our main example and analyzed the feature importance and interaction effects using EBM. The goal was to quantify the aging rate for this age group by identifying the most influential CpG sites and understanding their interactions. We can observe from Figure\ref{fig:age_20_30_features} that the top four CpG sites most important for this age group are cg16867657, cg10501210, cg22454769, and cg00329615. These sites have been identified in other studies \cite{Cpg1686&1050,Cpg,Cpg2245}, further validating the reliability of our model. As depicted in Figure \ref{fig:age_20_29_cpg_top1_2}, the methylation levels at cg16867657 range from 0.4 to 0.55, showing a positive linear relationship with their contribution to the model. Additionally, Figure \ref{fig:age_20_29_cpg_interaction} illustrates a strong positive contribution when both cg00573770 and cg16867657 are approximately 0.55, as well as when cg10501210 is around 0.5 and cg16867657 is approximately 0.6.

\section{Conclusion}

In this study, we introduced a novel two-phase approach for DNA methylation age prediction, specifically addressing the challenges of Epigenetic Correlation Drift (ECD) and Heterogeneity Among CpGs (HAC). By leveraging the Facebook AI Similarity Search (FAISS) for clustering and the Explainable Boosting Machine (EBM) for targeted, age-specific predictions, Proposed approach not only enhances the accuracy of age estimations but also improves interpretability. Specifically, the introduction of age group-specific feature selection allows for a more refined understanding of the methylation patterns associated with different stages of aging.

Our experimental results underscore the advantages of iTARGET-decade over traditional epigenetic clocks and classic linear regression models, as evidenced by lower MAE and RMSE across a broad age range. Furthermore, the decade-specific analysis of CpG site importance and interaction effects provides deeper insights into the biological aging process, revealing distinct patterns that correlate with different aging rates across the lifespan. Interactions between CpGs can be automatically detected using our method, providing an index for the next step in exploring the biology of synergistic interactions between CpGs and genes.

%\footnote{Acknowledgments: F.S. was Supported by a UKRI Future Leaders Fellowship, grant number MR/T043571/1.}

%\section*{Acknowledgments}

%\else
%  % regular IEEE prefers the singular form
\section*{Acknowledgments}

F.S. was supported by a UKRI Future Leaders Fellowship, grant number [MR/T043571/1].
%The authors would like to thank...


\begin{thebibliography}{10}

\bibitem{AgingClocksChallenges}
Christopher~G Bell, Robert Lowe, Peter~D Adams, Andrea~A Baccarelli, Stephan Beck, Jordana~T Bell, Brock~C Christensen, Vadim~N Gladyshev, Bastiaan~T Heijmans, Steve Horvath, et~al.
\newblock Dna methylation aging clocks: challenges and recommendations.
\newblock {\em Genome biology}, 20:1--24, 2019.

\bibitem{healthyaging}
Meaghan~J Jones, Sarah~J Goodman, and Michael~S Kobor.
\newblock Dna methylation and healthy human aging.
\newblock {\em Aging cell}, 14(6):924--932, 2015.

\bibitem{okada2023NonLinear}
Daigo Okada, Jian~Hao Cheng, Cheng Zheng, Tatsuro Kumaki, and Ryo Yamada.
\newblock Data-driven identification and classification of nonlinear aging patterns reveals the landscape of associations between dna methylation and aging.
\newblock {\em Human Genomics}, 17(1):8, 2023.

\bibitem{carlsen2023nonlinear2}
Laura Carlsen, Olivia Holl{\"a}nder, Moritz~Fabian Danzer, Marielle Vennemann, and Christa Augustin.
\newblock Dna methylation-based age estimation for adults and minors: considering sex-specific differences and non-linear correlations.
\newblock {\em International Journal of Legal Medicine}, 137(3):635--643, 2023.

\bibitem{NonlinearAging}
Xiaotao Shen, Chuchu Wang, Xin Zhou, Wenyu Zhou, Daniel Hornburg, Si~Wu, and Michael~P. Snyder.
\newblock Nonlinear dynamics of multi-omics profiles during human aging.
\newblock {\em Nature Aging}, 2024.
\newblock Shen2024.

\bibitem{belsky2022DunedinPACE}
Daniel~W Belsky, Avshalom Caspi, David~L Corcoran, Karen Sugden, Richie Poulton, Louise Arseneault, Andrea Baccarelli, Kartik Chamarti, Xu~Gao, Eilis Hannon, et~al.
\newblock Dunedinpace, a dna methylation biomarker of the pace of aging.
\newblock {\em Elife}, 11:e73420, 2022.

\bibitem{hannum2013genome}
Gregory Hannum, Justin Guinney, Ling Zhao, LI~Zhang, Guy Hughes, SriniVas Sadda, Brandy Klotzle, Marina Bibikova, Jian-Bing Fan, Yuan Gao, et~al.
\newblock Genome-wide methylation profiles reveal quantitative views of human aging rates.
\newblock {\em Molecular cell}, 49(2):359--367, 2013.

\bibitem{horvath2epigenetic2018}
Steve Horvath, Junko Oshima, George~M Martin, Ake~T Lu, Austin Quach, Howard Cohen, Sarah Felton, Mieko Matsuyama, Donna Lowe, Sylwia Kabacik, et~al.
\newblock Epigenetic clock for skin and blood cells applied to hutchinson gilford progeria syndrome and ex vivo studies.
\newblock {\em Aging (Albany NY)}, 10(7):1758, 2018.

\bibitem{HRSInCHPhenoAge}
Albert~T Higgins-Chen, Kyra~L Thrush, Yunzhang Wang, Christopher~J Minteer, Pei-Lun Kuo, Meng Wang, Peter Niimi, Gabriel Sturm, Jue Lin, Ann~Zenobia Moore, et~al.
\newblock A computational solution for bolstering reliability of epigenetic clocks: Implications for clinical trials and longitudinal tracking.
\newblock {\em Nature aging}, 2(7):644--661, 2022.

\bibitem{varshavsky2023accurate}
Miri Varshavsky, Gil Harari, Benjamin Glaser, Yuval Dor, Ruth Shemer, and Tommy Kaplan.
\newblock Accurate age prediction from blood using a small set of dna methylation sites and a cohort-based machine learning algorithm.
\newblock {\em Cell Reports Methods}, 3(9), 2023.

\bibitem{lin2016dna}
Qiong Lin, Carola~I Weidner, Ivan~G Costa, Riccardo~E Marioni, Marcelo~RP Ferreira, Ian~J Deary, and Wolfgang Wagner.
\newblock Dna methylation levels at individual age-associated cpg sites can be indicative for life expectancy.
\newblock {\em Aging (Albany NY)}, 8(2):394, 2016.

\bibitem{PhenoAge}
Morgan~E Levine, Ake~T Lu, Austin Quach, Brian~H Chen, Themistocles~L Assimes, Stefania Bandinelli, Lifang Hou, Andrea~A Baccarelli, James~D Stewart, Yun Li, et~al.
\newblock An epigenetic biomarker of aging for lifespan and healthspan.
\newblock {\em Aging (albany NY)}, 10(4):573, 2018.

\bibitem{YingCausAgeDamAgeAdaptAge}
Kejun Ying, Hanna Liu, Andrei~E Tarkhov, Marie~C Sadler, Ake~T Lu, Mahdi Moqri, Steve Horvath, Zolt{\'a}n Kutalik, Xia Shen, and Vadim~N Gladyshev.
\newblock Causality-enriched epigenetic age uncouples damage and adaptation.
\newblock {\em Nature aging}, 4(2):231--246, 2024.

\bibitem{DL2022pan}
Lucas~Paulo de~Lima~Camillo, Louis~R Lapierre, and Ritambhara Singh.
\newblock A pan-tissue dna-methylation epigenetic clock based on deep learning.
\newblock {\em npj Aging}, 8(1):4, 2022.

\bibitem{galkin2021deepLmage}
Fedor Galkin, Polina Mamoshina, Kirill Kochetov, Denis Sidorenko, and Alex Zhavoronkov.
\newblock Deepmage: a methylation aging clock developed with deep learning.
\newblock {\em Aging and disease}, 12(5):1252, 2021.

\bibitem{deeplNN2021age}
Lechuan Li, Chonghao Zhang, Shiyu Liu, Hannah Guan, and Yu~Zhang.
\newblock Age prediction by dna methylation in neural networks.
\newblock {\em IEEE/ACM Transactions on Computational Biology and Bioinformatics}, 19(3):1393--1402, 2021.

\bibitem{levy2020methylnetdeepl}
Joshua~J Levy, Alexander~J Titus, Curtis~L Petersen, Youdinghuan Chen, Lucas~A Salas, and Brock~C Christensen.
\newblock Methylnet: an automated and modular deep learning approach for dna methylation analysis.
\newblock {\em BMC bioinformatics}, 21:1--15, 2020.

\bibitem{vidaki2017dnaDeepl}
Athina Vidaki, David Ballard, Anastasia Aliferi, Thomas~H Miller, Leon~P Barron, and Denise~Syndercombe Court.
\newblock Dna methylation-based forensic age prediction using artificial neural networks and next generation sequencing.
\newblock {\em Forensic Science International: Genetics}, 28:225--236, 2017.

\bibitem{qiu2023EAIIMLAgeClock}
Wei Qiu, Hugh Chen, Matt Kaeberlein, and Su-In Lee.
\newblock Explainable biological age (enabl age): an artificial intelligence framework for interpretable biological age.
\newblock {\em The Lancet Healthy Longevity}, 4(12):e711--e723, 2023.

\bibitem{douze2024faiss}
Matthijs Douze, Alexandr Guzhva, Chengqi Deng, Jeff Johnson, Gergely Szilvasy, Pierre-Emmanuel Mazar{\'e}, Maria Lomeli, Lucas Hosseini, and Herv{\'e} J{\'e}gou.
\newblock The faiss library.
\newblock {\em arXiv preprint arXiv:2401.08281}, 2024.

\bibitem{EBM2022interpretability}
Zijie~J Wang, Alex Kale, Harsha Nori, Peter Stella, Mark~E Nunnally, Duen~Horng Chau, Mihaela Vorvoreanu, Jennifer~Wortman Vaughan, and Rich Caruana.
\newblock Interpretability, then what? editing machine learning models to reflect human knowledge and values.
\newblock {\em arXiv preprint arXiv:2206.15465}, 2022.

\bibitem{lou2013EBM}
Yin Lou, Rich Caruana, Johannes Gehrke, and Giles Hooker.
\newblock Accurate intelligible models with pairwise interactions.
\newblock In {\em Proceedings of the 19th ACM SIGKDD international conference on Knowledge discovery and data mining}, pages 623--631. ACM, 2013.

\bibitem{caruana2015EBM}
Rich Caruana, Yin Lou, Johannes Gehrke, Paul Koch, Marc Sturm, and Noemie Elhadad.
\newblock Intelligible models for healthcare: Predicting pneumonia risk and hospital 30-day readmission.
\newblock In {\em Proceedings of the 21th ACM SIGKDD International Conference on Knowledge Discovery and Data Mining}, pages 1721--1730. ACM, 2015.

\bibitem{lehallier2019undulating}
Benoit Lehallier, David Gate, Nicholas Schaum, Tibor Nanasi, Song~Eun Lee, Hanadie Yousef, Patricia Moran~Losada, Daniela Berdnik, Andreas Keller, Joe Verghese, et~al.
\newblock Undulating changes in human plasma proteome profiles across the lifespan.
\newblock {\em Nature medicine}, 25(12):1843--1850, 2019.

\bibitem{horvath2013dna}
Steve Horvath.
\newblock Dna methylation age of human tissues and cell types.
\newblock {\em Genome biology}, 14:1--20, 2013.

\bibitem{ElasticNet}
Hui Zou and Trevor Hastie.
\newblock Regularization and variable selection via the elastic net.
\newblock {\em Journal of the Royal Statistical Society Series B: Statistical Methodology}, 67(2):301--320, 2005.

\bibitem{Cpg1686&1050}
Ziwei Ye, Lirong Jiang, Mengyao Zhao, Jing Liu, Hao Dai, Yiping Hou, and Zheng Wang.
\newblock Epigenome-wide screening of cpg markers to develop a multiplex methylation snapshot assay for age prediction.
\newblock {\em Legal Medicine}, 59:102115, 2022.

\bibitem{Cpg}
Anastasia Aliferi, Sudha Sundaram, David Ballard, Ana Freire-Aradas, Christopher Phillips, Maria~Victoria Lareu, and Denise~Syndercombe Court.
\newblock Combining current knowledge on dna methylation-based age estimation towards the development of a superior forensic dna intelligence tool.
\newblock {\em Forensic Science International: Genetics}, 57:102637, 2022.

\bibitem{Cpg2245}
VA~Lemesh, VN~Kipen, MV~Bahdanava, AA~Burakova, AG~Bulgak, AV~Bayda, SA~Bruskin, OV~Zotova, and OI~Dobysh.
\newblock Determination of human chronological age from biological samples based on the analysis of methylation of cpg dinucleotides.
\newblock {\em Russian Journal of Genetics}, 57:1389--1397, 2021.

\end{thebibliography}
\end{document}